
\input harvmac.tex
\Title{CTP/TAMU-89/91}{String and Fivebrane Solitons:
Singular or Non-singular?\footnote{$^\dagger$}
{Work supported in part by NSF grant PHY-9106593.}}

\centerline{M.~J.~Duff,
R.~R.~ Khuri\footnote{$^*$}{Supported by a World Laboratory Fellowship.}
and J.~X.~Lu\footnote{$^{**}$}{Supported by a World Laboratory Scholarship.}}
\bigskip\centerline{Center for Theoretical Physics}
\centerline{Texas A\&M University}\centerline{College Station, TX 77843}

\vskip .3in
We ask whether the recently discovered superstring and superfivebrane solutions
of $D=10$ supergravity admit the interpretation of non-singular solitons
even though, in the absence of Yang-Mills fields, they exhibit curvature
singularities at the origin. We answer the question using a
test probe/source approach, and find that the nature of the singularity
is probe-dependent. If the test probe and source are both superstrings or both
superfivebranes, one falls into the other in a finite proper time and the
singularity is real, whereas if one is a superstring and the other a
superfivebrane it takes an infinite proper time (the force is repulsive!)
and the singularity is harmless.
Black strings and fivebranes, on the other hand, always display real
singularities.

\Date{11/91}
\lref\ccrk{C.~G.~Callan and R.~R.~Khuri, Phys. Lett. {\bf B261} (1991) 363.}

\lref\rkthree{R.~R.~Khuri, {\it Manton Scattering of String Solitons}
PUPT-1270, August 1991.}

\lref\dghrr{A.~Dabholkar, G.~Gibbons, J.~A.~Harvey and F.~Ruiz Ruiz,
Nucl. Phys. {\bf B340} (1990) 33.}

\lref\dfluone{M.~J.~Duff and J.~X.~Lu, Nucl. Phys. {\bf B354} (1991) 141.}

\lref\dflutwo{M.~J.~Duff and J.~X.~Lu, Nucl. Phys. {\bf B354} (1991) 129.}

\lref\dfluthree{M.~J.~Duff and J.~X.~Lu, Phys. Rev. Lett. {\bf 66}
(1991) 1402.}

\lref\dflufour{M.~J.~Duff and J.~X.~Lu, Nucl. Phys. {\bf B357} (1991)
534.}

\lref\dflufive{M.~J.~Duff and J.~X.~Lu, {\it A Duality Between Strings and
Fivebranes}, CTP-TAMU-28/91 (to appear in Class. Quant. Gravity).}

\lref\dflusix{M.~J.~Duff and J.~X.~Lu, {\it The Self-Dual Type II-B
Superthreebrane}, CTP-TAMU-29/91 (to appear in Phys. Lett. {\bf B}).}

\lref\chsone{C.~G.~Callan, J.~A.~Harvey and A.~Strominger, Nucl. Phys.
{\bf B359} (1991) 611.}

\lref\chstwo{C.~G.~Callan, J.~A.~Harvey and A.~Strominger {\it
Worldbrane Action for String Solitons} PUPT-1244.}

\lref\strom{A.~Strominger, Nucl. Phys. {\bf B343} (1990) 167.}

\lref\dfst{M.~J.~Duff and K.~S.~Stelle, Phys. Lett. {\bf B253} (1991)
113.}

\lref\host{G.~T.~Horowitz and A.~Strominger, Nucl. Phys. {\bf B360}
(1991) 197.}

\lref\hlp{J.~Hughes, J.~Liu and J.~Polchinski, Phys. Lett. {\bf B180}
(1986).}

\lref\town{P.~K.~Townsend, Phys. Lett. {\bf B202} (1988) 53.}

\lref\duff{M.~J.~Duff, Class. Quant. Grav. {\bf 5} (1988).}

\newsec{Introduction}
Dabholkar {\it et al.}\dghrr\ have shown that the field equations of $D=10$
$N=1$ supergravity with a $3$-form field strength admit as a solution
an infinite string which breaks half the spacetime supersymmetries.
Similarly, Duff and Lu\dfluone\ have shown
that the field equations of the dual version of $D=10$ $N=1$ supergravity
with a $7$-form field strength admit as a solution an infinite fivebrane
which also breaks half the spacetime supersymmetries. The same
string and fivebrane configurations, which are reviewed in section 2,
also solve both the Type IIA and Type IIB supergravity
equations\refs{\dghrr,\chstwo}.
Both solutions possess a nonvanishing ``electric" Noether charge, conserved
as a consequence of the equations of motion. Both saturate a Bogomol'nyi
bound between the mass per unit length or $5$-volume and the Noether charge.
Moreover, both obey a ``no static force" condition between parallel strings
or fivebranes of the same orientation, the gravitational attraction of the
graviton and dilaton being exactly cancelled by a repulsion due to the
antisymmetric tensor. As such, the solutions can be generalized to
multi-string or multi-fivebrane solutions.

In \ccrk, Callan and Khuri employed a test string approximation to the
dynamics of the string solutions with the result that a test string moving
in the background of a parallel Dabholkar source string with the
same orientation experienced zero dynamical force to lowest order in the
velocity of the test string. This result was found to be consistent with
a Veneziano amplitude calculation for the scattering of infinitely long
macroscopic strings, and was used to provide dynamical evidence for the
identification of the Dabholkar solution with
the fundamental string. An analogous calculation for the fivebrane solutions
of Duff and Lu also yielded a vanishing lowest order dynamical force
on a test fivebrane in the background of a source fivebrane with identical
orientation.

One source of confusion, however, is whether these solutions admit the
interpretation of non-singular ``solitons" which, after all, provided the
original motivation for supermembranes\refs{\hlp,\town}. At first sight the
answer is no because both the string solution and the fivebrane solution
were obtained by coupling supergravity to the corresponding $\sigma$-models
and hence displayed $\delta$-function singularities at the location of the
extended object source, $r=0$. Moreover, in both cases the curvature calculated
from the $\sigma$-model metric blows up at $r=0$.
In these respects they seem to differ from the
heterotic fivebrane solution of Strominger\strom\ and the heterotic
string solution of Duff and Lu\dfluthree\ where the singularities are
smeared out by the presence of the Yang-Mills fields. These latter solutions
are
genuine ``solitons", possessing a non-vanishing ``magnetic" topological
charge, identically conserved as a consequence of the Bianchi identities.
Indeed, the singular Dabholkar {\it et al.} string is obtained from the
nonsingular Duff-Lu string by shrinking to zero the size of the Yang-Mills
instanton that lives in the $8$ dimensions transverse to the string\dfluthree.
Similarly, the singular Duff-Lu fivebrane is obtained from the non-singular
Strominger fivebrane by shrinking to zero the size of the Yang-Mills
instanton that lives in the $4$ dimensions transverse to the fivebrane\dfluone.
This correspondence is best understood when viewed from the
point of view of the {\it dual} theory which, by interchanging
$3$-form and $7$-form, interchanges field equations with Bianchi identities
and hence interchanges Noether electric charges with topological magnetic
charges\refs{\dflutwo,\dflufive}.

In \chsone, Callan, Harvey and Strominger argue that, from this point
of view, even the Duff-Lu fivebrane admits the interpretation
of a {\it non-singular} soliton solution of the {\it source-free}
$3$-form equations. They claim that the singularity at $r=0$ is only a
coordinate singularity.
In \host, Horowitz and Strominger point out that both solutions are
the extremal mass$=$charge limit of more general solutions with event horizons.
They state that for the extremal black fivebrane, both the event horizon
and the singularity disappear, but that the event horizon becomes
singular for the extremal black string.

In this paper, we attempt to clarify the nature of these singularities
by asking a physical question: Does a test-probe fall into the source in a
finite proper time, as measured by its own clock, in
which case the singularity is real, or in an infinite proper time, in
which case the singularity is harmless? We find that the answer is
probe-dependent. We show
in section 3 that when test probe and source are both strings or both
fivebranes the singularity is real. By contrast, in section 4, we show
that if one is a string and the other a fivebrane, the singularity is
harmless. In section 5, we strengthen this idea by noting that {\it both}
curvature singularities disappear if the metrics are re-expressed in the
{\it dual} variables.
Thus the singularity structure is symmetric between strings
and fivebranes, in accordance with the idea of string/fivebrane
duality\refs{\duff,\strom,\dflutwo,\dflufive}.

In section 6 we consider the recently discovered $D=10$ self-dual
type IIB superthreebrane\dflusix. We find that a test threebrane takes a finite
amount of proper time to reach a source threebrane, but in this case there
never was a curvature singularity in the first place. In section 7 a
similar result is seen to hold for the $D=11$ supermembrane which is related
to the $D=10$ superstring by simultaneous dimensional reduction\dfst.
Finally in section 8, we note that black p-branes, as opposed to the super
p-branes, always display real singularities.

\newsec{The Elementary String and the Elementary Fivebrane}

The elementary string solution of the three-form version of $D=10$
supergravity found by Dabholkar {\it et al.}\dghrr\ corresponds to supergravity
coupled to the string $\sigma$-model. The action for the supergravity fields
$(g_{MN},B_{MN},\phi)$ is given by
\eqn\smf{S({\rm string}) = {1\over 2\kappa^2}\int d^{10} x \sqrt{-g}\bigg(R -
{1\over 2}(\partial\phi)^2 - {1\over 2.3!}e^{-\phi}H^2\bigg)}
where $H=dB$ and the string $\sigma$-model action is given by
\eqn\stwo{S_2 = - T_2 \int d^2 \xi \bigg({1\over 2}\sqrt{-\gamma}\gamma^{ij}
\partial_i X^{M} \partial_j X^{N} g_{MN} e^{\phi/2}
+ {1\over 2}\epsilon^{ij}
\partial_i X^{M} \partial_j X^{N} B_{MN}\bigg)}
where $M,N=0,1,...,9$.
We have denoted the string tension by $T_2 = 1/\alpha'$.
The solution to the equations of motion of the combined action
$S({\rm string})+S_2$ is given by
\eqn\dabsol{\eqalign{e^{-2\phi}&=1+{k_2\over r^6},\cr
 ds^2&=e^{3\phi/2}\eta_{\mu\nu}dx^\mu dx^\nu+e^{-\phi/2}\delta_{mn}dx^mdx^n\cr
 B_{01}&=-e^{2\phi}\cr}}
where $\mu,\nu=0,1$ and $m,n=2,3,...,9$
and where $r$ is the radial coordinate for the eight-dimensional space
transverse to the two-dimensional worldsheet. For convenience,
we have taken the vev of the dilaton
$\phi_0=0$. The constant $k_2$ is given by
$k_2={\kappa^2 T_2\over 3\Omega_7}$ where $\Omega_n$
is the volume of the unit $n$-sphere. Note that the string couples to a metric
related to the canonical metric of \smf\ by
\eqn\scansig{g_{MN}({\rm string\ \sigma-model})=
e^{\phi/2}g_{MN}({\rm canonical}).}
We may verify that this metric exhibits a curvature singularity at $r=0$
by computing the scalar curvature
\eqn\stcurv{R_{\rm string}({\rm string\ \sigma-model})\sim -r^{-2}.}

The elementary fivebrane solution of the seven-form version of $D=10$
supergravity found by Duff and Lu\dfluone\ corresponds to supergravity
coupled to the fivebrane $\sigma$-model. The action for the supergravity fields
$(g_{MN},A_{MNPQRS},\phi)$ is given by
\eqn\fmf{S({\rm fivebrane}) = {1\over 2\kappa^2}~\int~d^{10} x
\sqrt{-g}\bigg(R - {1\over 2}
(\partial \phi)^2 - {1\over 2.7!}~e^{\phi} K^2\bigg)}
where $K=dA$ and the fivebrane $\sigma$-model action is given
by\refs{\dfluone,\dflutwo}
\eqn\ssix{\eqalign{S_6 =&- T_6 \int d^6 \xi \bigg({1\over 2} \sqrt{-\gamma}
\gamma^{ij}
\partial_i X^M \partial_j X^N g_{MN} e^{-\phi/6}
- 2 \sqrt{-\gamma}\cr
&+ {1\over 6!} \epsilon^{mnpqrs} \partial_m X^M \partial_n X^N
\partial_p X^P \partial_q X^Q \partial_r X^R \partial_s
X^S A_{MNPQRS}\bigg)\cr}}
where we have denoted the fivebrane tension by $T_6$.
The solution to the equations of motion of the combined action
$S({\rm fivebrane})+S_6$ is given by
\eqn\dflusol{\eqalign{e^{2\phi}&=1+{k_6\over r^2}\cr
 ds^2&=e^{-\phi/2}\eta_{\mu\nu}dx^\mu dx^\nu+e^{3\phi/2}\delta_{mn}dx^mdx^n\cr
 A_{012345}&=-e^{-2\phi}\cr}}
where $\mu,\nu=0,1,2,...,5$ and $m,n=6,7,8,9$
and where $r$ is the radial coordinate for the four-dimensional
space transverse to the six-dimensional worldvolume.
Again we set $\phi_0=0$. The constant $k_6$ is given by $k_6={\kappa^2T_6\over
\Omega_3}$. Note that the fivebrane couples to a metric related to the
canonical
metric of \fmf\ by
\eqn\fcansig{g_{MN}({\rm fivebrane\ \sigma-model})=
e^{-\phi/6}g_{MN}({\rm canonical}).}
We may verify that this metric exhibits a curvature singularity at $r=0$
by computing the scalar curvature
\eqn\fbcurv{R_{\rm fivebrane}({\rm fivebrane\ \sigma-model})\sim
-(k_6r)^{-2/3}.}
(Incidentally, $R_{\rm string}$ and $R_{\rm fivebrane}$ both blow up as
$r^{-1/2}$ in canonical variables.)

\newsec{String-String and Fivebrane-Fivebrane Radial Trajectories}

Let us consider the trajectory of a test string falling radially into
a source string, oriented along $x^1=\xi^1$. For simplicity, let the test
string lie either parallel or antiparallel to the source string. If we
eliminate $\gamma_{ij}$ from \stwo\ and substitute \dabsol, we find that
the Lagrangian governing the dynamics of the test string is given by
\eqn\stradlag{{\cal L}_2=-T_2 e^{2\phi}\left(\sqrt{\dot t^2-\dot r^2
e^{-2\phi}}\mp \dot t\right)}
where the minus (plus) sign corresponds to the parallel (antiparallel)
configuration. The time derivative is with respect to $\xi^0$ which
we choose to be the proper time $\tau$ measured by a clock at rest in
the frame of the test string. From \dabsol\ and \scansig\ this is given
by
\eqn\stlinel{d\tau^2=-e^{\phi/2}ds^2=e^{2\phi}dt^2-dr^2.}
Thus the calculation has been reduced to a one-dimensional problem and
the dynamics of \stradlag\ is similar to that of a point particle whose
mass is equal or opposite to its electric charge. Since there is
no explicit time-dependence in ${\cal L}_2$, we have the following constant of
the motion
\eqn\estst{{\partial{\cal L}_2\over \partial \dot t}=-T_2 e^{2\phi}
\left({\dot t\over \sqrt{\dot t^2-\dot r^2 e^{-2\phi}}}\mp 1\right)=
-T_2(E\mp 1).}
$E$ is the constant energy per unit mass of the motion and is determined
from the intial conditions. Note that for the parallel strings case, we
recover the zero static force result by noting that if $\dot r=0$ initially,
then $E=1$ and $\dot r=0$ everywhere. We also recover
the vanishing leading order (in the velocity) dynamic force result found in
\ccrk.
{}From \estst\ we obtain an expression for the coordinate velocity
\eqn\ssv{\left({dr\over dt}\right)^2={(E\mp 1)^2 e^{-2\phi}\pm 2(E\mp 1)
\over \left((E\mp 1) e^{-2\phi}\pm 1\right)^2}.}
We now wish to relate the radial position to the proper time.
Combining \ssv\ and \stlinel\ we obtain
\eqn\sspv{\left({dr\over d\tau}\right)^2=e^{-2\phi}\left((E\mp 1)^2
e^{-2\phi}\pm
2(E\mp 1)\right)}
for the proper velocity in terms of the radial position. The
acceleration can be obtained by differentiating \sspv\ with respect to
$\tau$ and replacing \sspv\ in the resulting expression. The acceleration
written in terms of the position is independent of the sign of the velocity
and is given by
\eqn\sspa{{d^2r\over d\tau^2}=-{6k_2(E\mp 1)^2\over r^7}\left(1+{k_2\over r^6}
\pm (E\mp 1)^{-1}\right).}
For parallel strings, the force is always attractive when initially
$\dot r\neq 0$. For antiparallel strings,
the acceleration is always inward, and the test string does
indeed fall towards the source string. We may thus choose the negative
sign for the square roots in \ssv\ and \sspv. To calculate the proper time
taken for the test string to reach the source string, we rewrite \sspv\
and integrate
\eqn\sspt{\tau_0=\int_0^{\tau_0}d\tau=\int_0^{r_0}{dr\over
\sqrt{e^{-2\phi}\left((E\mp 1)^2e^{-2\phi}\pm 2(E\mp 1)\right)}}.}
On using the expression for $\phi$ in \dabsol, we
note that $\tau_0$ is finite.
Thus the test string falls into the source string
in a finite amount of time, and the singularity is real. In particular,
let us focus on the case where the test string is antiparallel to the source
string. If $\dot r=0$ at $r=r_0$, then
\eqn\estinit{E+1=2e^{2\phi(r_0)}={2\over 1+k_2/r_0^6}.}
Let $x\equiv r/r_0$, then $\tau_0$ can be written as
\eqn\sspttwo{\tau_0={e^{-2\phi(r_0)}r_0^4\over 2\sqrt{k_2}}
\int_0^1 {dx x^6\over \sqrt{(x^6+k_2/r_0^6)(1-x^6)}}.}
For large $r_0$, we find that $\tau_0\sim k_2^{-1/2}$. Since the mass per
unit length of the string is given by $M_2=T_2$\dghrr, this means that
$\tau_0\sim M_2^{-1/2}$ which is the same dependence of the time on the mass
for an observer falling into a Schwarzschild black hole. Just as for the
black hole case, moreover, it is easy to see from \sspv\ and \sspa\ that
the proper velocity and acceleration both tend to infinity as the
test string approaches the singularity.
To further strengthen the analogy with a black hole-type singularity,
one can calculate the elapsed distant observer time for the fall. In this case
one can easily show that $t_0\to\infty$, $dr/dt\to 0$ and $d^2r/dt^2\to 0$
as the test string approaches the singularity. In other words, the distant
observer never sees the test string reach the singularity. In this case,
the event horizon is at the singularity.

We shall now repeat the above calculation for a test-fivebrane falling
radially into a source fivebrane oriented along $x^a=\xi^a$, ($a=1,...,5$).
Again, we let the test fivebrane lie either parallel or antiparallel
to the source fivebrane, i.e. with the same or opposite orientation. If
we eliminate $\gamma_{ij}$ from \ssix\ and \dflusol, we find that the
Lagrangian governing the dynamics of the test fivebrane is given by
\eqn\fbradlag{{\cal L}_6=-T_6 e^{-2\phi}\left(\sqrt{\dot t^2-\dot r^2
e^{2\phi}}
\mp \dot t\right)}
where the minus (plus) sign corresponds to the parallel (antiparallel)
configuration. The time derivative is with respect to $\xi^0$, which
we choose to be the proper time $\tau$ measured by a clock at rest in
the frame of the test fivebrane. From \dflusol\ and \fcansig\ this is given
by
\eqn\fblinel{d\tau^2=-e^{-\phi/6}ds^2=e^{-2\phi/3}\left(dt^2-e^{2\phi}dr^2
\right).}
This time the Euler-Lagrange equations yield the following constant of the
motion
\eqn\efbfb{{\partial{\cal L}_6\over \partial \dot t}=-T_6 e^{-2\phi}
\left({\dot t\over \sqrt{\dot t^2-\dot r^2 e^{2\phi}}}\mp 1\right)=
-T_6(E\mp 1).}
{}From \efbfb\ we obtain an expression for the coordinate velocity
\eqn\ffv{\left({dr\over dt}\right)^2={(E\mp 1)^2 e^{2\phi}\pm 2(E\mp 1)\over
\left((E\mp 1) e^{2\phi}\pm 1\right)^2}.}
Combining \ffv\ and \fblinel\ we obtain
\eqn\ffpv{\left({dr\over d\tau}\right)^2=e^{2\phi/3}
\left((E\mp 1)^2 e^{2\phi}\pm 2(E\mp 1)\right)}
for the proper velocity in terms of the radial position. The
acceleration can be obtained by differentiating \ffpv\ with respect to
$\tau$ and replacing \ffpv\ in the resulting expression. The acceleration
written in terms of the position is independent of the sign of the velocity
as in the string-string case and is again attractive, so the test fivebrane
does
indeed fall towards the source fivebrane. To calculate the proper time
taken for the test fivebrane to reach the source fivebrane we rewrite \ffpv\
and integrate
\eqn\ffpt{\tau_0=\int_0^{\tau_0}d\tau=\int_0^{r_0}{dr\over
\sqrt{e^{2\phi/3}\left((E\mp 1)^2e^{2\phi}\pm 2(E\mp 1)\right)}}.}
On using the expression for $\phi$ in \dflusol, we note
again that $\tau_0$ is manifestly finite.
Thus the test fivebrane falls into the source fivebrane
in a finite amount of time, and the singularity is real. In the antiparallel
case, the dependence of the time on the mass of a source for the test fivebrane
initially at rest and for large initial separation is again of the form
$\tau_0\sim k_6^{-1/2}$. Since the mass per unit $5$-volume of the fivebrane
is given by $M_6=T_6$\dfluone, this means $\tau_0\sim M_6^{-1/2}$ as for the
string. It is easy to see
that the proper velocity and acceleration both $\to\infty$ as the
test fivebrane approaches the singularity
and that $t_0\to\infty$, $dr/dt\to 0$ and $d^2r/dt^2\to 0$.
Once more, the event horizon is located at the singularity.

\newsec{String-Fivebrane and Fivebrane-String Radial Trajectories}
An entirely different state of affairs holds for a test string moving
in the background of a source fivebrane or, by duality, a test fivebrane
moving in the background of a source string. In this case, the test probe
takes an infinite amount of proper time to reach the source.

The actions $S({\rm string})$ and $S({\rm fivebrane})$ become dual to each
other if we make the identification\refs{\dflutwo,\dflufive}
\eqn\duality{K=e^{-\phi}{}{}^*H.}
In this case, the field equation for $H$ becomes the Bianchi identity for
$K$ and vice-versa. We shall make use of this duality in discussing how
a test string behaves in the field of a source fivebrane, and how a test
fivebrane behaves in the field of a string.

First we consider the trajectory of a test string falling radially into a
source fivebrane, oriented along $x^a=\xi^a$ ($a=1,2,...,5$). Let the test
string lie either parallel or antiparallel to one of the fivebrane directions,
say $x^1$. From \dflusol, the only nonvanishing components of $K$
are of the form $K_{012345m}$, where the directions $m=6,7,8,9$
are transverse to the
fivebrane. By dualizing, we see that the only nonzero components of
$H=dB$ are $H_{pqr}(r)$, where again, $p,q,r=6,7,8,9$. It follows that
the only nonzero components of $B_{MN}$ occur when $M,N=6,7,8,9$. It then
follows that the WZW term $\epsilon^{ij}\partial_iX^M\partial_jX^NB_{MN}$
vanishes. Substituting \dflusol\ in \stwo, we find that the test string
Lagrangian reduces to
\eqn\stfblag{{\cal L}_2=-T_2\sqrt{\dot t^2-e^{2\phi}\dot r^2}}
for purely radial motion. From \dflusol\ and \scansig, the proper time
is given by
\eqn\stfblinel{d\tau^2=-e^{\phi/2}ds^2=dt^2-e^{2\phi}dr^2.}
Again we have a constant of the motion
\eqn\stfbenergy{{\partial{\cal L}_2\over \partial\dot t}=-T_2{\dot t\over
\sqrt{\dot t^2-e^{2\phi}\dot r^2}}=-T_2 E.}
Note that $E=1$ corresponds to a zero static force.
We invert \stfbenergy\ to obtain the coordinate velocity
\eqn\sfv{\left({dr\over dt}\right)^2=e^{-2\phi}\left(1-1/E^2\right).}
Combining \stfblinel\ and \stfbenergy, we obtain the proper velocity
\eqn\sfpv{\left({dr\over d\tau}\right)^2=\left(E^2-1\right)e^{-2\phi}.}
The acceleration is given by
\eqn\sfpa{{d^2r\over d\tau^2}={k_6(E^2-1)e^{-4\phi}\over r^3}.}
Note that the acceleration is repulsive in this case.
In both the $r\to 0$ and $r\to\infty$ limits, the acceleration vanishes
(an asymptotic freedom of some sort). We assume that the string is directed
towards the fivebrane initially. The time taken for the fall from
an initial position $r_0$
\eqn\sfpt{\tau_0={1\over\sqrt{E^2-1}}\int_0^{r_0} e^\phi dr}
diverges logarithmically with $r$. Therefore it takes the string an
infinite amount of proper time to reach the singularity. In other words,
the string never sees the singularity, and as far as the string is concerned,
the singularity is not real. In this sense, the Duff-Lu fivebrane solution
is non-singular.

An analogous calculation for a test fivebrane falling
towards a source string shows that the string is nonsingular
as a source for fivebranes. For a test fivebrane with one of its
spatial directions parallel to the string, the WZW term again vanishes, as
in the above case. In this case, substituting \dabsol\ into \ssix,
the Lagrangian reduces to
\eqn\fbstlag{{\cal L}_6=-T_6\sqrt{\dot t^2-e^{-2\phi}\dot r^2}}
for purely radial motion. From \dabsol\ and \fcansig, the proper time is
given by
\eqn\fbstlinel{d\tau^2=-e^{-\phi/6}ds^2=e^{4\phi/3}\left(dt^2-e^{-2\phi}dr^2
\right).}
Again we have a constant of the motion
\eqn\fbstenergy{{\partial{\cal L}_6\over \partial\dot t}=-T_6{\dot t\over
\sqrt{\dot t^2-e^{2\phi}\dot r^2}}=-T_6 E.}
Again $E=1$ corresponds to a zero static force.
We invert \fbstenergy\ to obtain the coordinate velocity
\eqn\fsv{\left({dr\over dt}\right)^2=e^{2\phi}\left(1-1/E^2\right).}
Combining \fsv\ and \fbstlinel\ we obtain the proper velocity
\eqn\fspv{\left({dr\over d\tau}\right)^2=\left(E^2-1\right)e^{2\phi/3}.}
The acceleration is again found to be repulsive and is given by
\eqn\fspa{{d^2r\over d\tau^2}={k_2(E^2-1)e^{8\phi/3}\over r^7}.}
Again the acceleration vanishes in both the $r\to 0$ and $r\to\infty$ limits.
Now assume that the fivebrane is directed
towards the string initially. The time taken for the fall from
an initial position $r_0$ is given by
\eqn\fspt{\tau_0={1\over\sqrt{E^2-1}}\int_0^{r_0} e^{-{\phi\over 3}} dr}
and again diverges logarithmically with $r$. Therefore it takes the
fivebrane an infinite amount of proper time to reach the singularity. In
other words, the fivebrane never sees the singularity, and as far as the
fivebrane is concerned, the string singularity is not real. In this sense,
the Dabholkar {\it et al.} string solution is non-singular.

\newsec{Curvature Singularities Re-examined}

In section 3 we found that the string-string and fivebrane-fivebrane
singularities were real in accordance with the results of \stcurv\
and \fbcurv\ that both the string metric in string variables and the
fivebrane metric in fivebrane variables exhibited curvature singularities
at $r=0$. However, in section 4 we found that the string-fivebrane and
fivebrane-string singularities were harmless. This suggests that we should
reexamine the curvature singularities by recasting the string metric \dabsol\
in fivebrane variables \fcansig\ and the fivebrane metric \dflusol\
in string variables \scansig. It is instructive to employ polar coordinates.
In this case the string solution is
\eqn\dabfb{-d\tau^2=\left(1+{k_2\over r^6}\right)^{-2/3}\eta_{\mu\nu}
dx^\mu dx^\nu + \left(1+{k_2\over r^6}\right)^{1/3}\left(dr^2+r^2d\Omega_7^2
\right)}
and the fivebrane solution is
\eqn\dflust{-d\tau^2=\eta_{\mu\nu}
dx^\mu dx^\nu + \left(1+{k_6\over r^2}\right)\left(dr^2+r^2d\Omega_3^2\right).}
It is these metrics that provide the relevant proper time in \fbstlinel\
and \stfblinel. Remarkably, both are free of curvature singularities,
as may be seen by noting
that, as $r\to 0$, the radius of $S^7$ in \dabfb\ tends to the finite
value $k_2^{1/6}$ and the radius of $S^3$ in \dflust\ tends to the finite
value $k_6^{1/2}$. This is confirmed by a calculation of the scalar
curvatures. We find
\eqn\scurvtwo{R_{\rm string}({\rm fivebrane\ \sigma-model})\sim +k_2^{-1/3}}
\eqn\fcurvtwo{R_{\rm fivebrane}({\rm string\ \sigma-model})\sim +k_6^{-1}.}
Thus the singularity structure is entirely symmetric between strings and
fivebranes, in accordance with string/fivebrane duality. [Note that throughout
this paper we have employed the fivebrane $\sigma$-model metric of
\refs{\dfluone,\dflutwo}\ given in \fcansig, for which
$g_{MN}({\rm fivebrane\ \sigma-model})=e^{-2\phi/3}
g_{MN}({\rm string\ \sigma-model})$. Now any metric
$e^{-a\phi}g_{MN}({\rm string\ \sigma-model})$ will yield non-singular string
solutions and any metric $e^{+a\phi}g_{MN}({\rm fivebrane\ \sigma-model})$
will yield non-singular fivebrane solutions provided $a\geq 2/3$. In
particular, the choice $g^*_{MN}=e^{-2\phi}g_{MN}({\rm string\ \sigma-model})$
yields a non-singular string solution, as pointed out by Callan, Harvey
and Strominger\chstwo. However, only the choice $a=2/3$ enters into the
string-fivebrane and fivebrane-string calculations of section 4, and so in
this context we do not attach any physical significance to other choices.]

\newsec{The Self-Dual Threebrane}

The results of sections 2--5 seem to suggest a correlation between the
existence
of a curvature singularity and the proper time taken to reach $r=0$. Before
jumping to conclusions, however, it is instructive to examine the recently
discovered $D=10$ self-dual Type IIB superthreebrane\dflusix. The bosonic
equation of motion reads
\eqn\tmotion{\eqalign{&\partial_i\left(\sqrt{-\gamma}\gamma^{ij}\partial_j
X^N g_{MN}\right)-{1\over 2}\sqrt{-\gamma}\gamma^{ij}\partial_i X^N \partial_j
X^P \partial_M g_{NP}\cr &-{1\over 4}\epsilon^{ijkl}\partial_i X^N
\partial_j X^P \partial_k X^Q \partial_l X^R F_{MNPQR}=0.\cr}}
However, because the rank-five field strength is anti self-dual
\eqn\asdual{F=-{}^* F}
there is no covariant action $S_4$ akin to \stwo\ or \ssix. The solution
reads
\eqn\tbrane{\eqalign{e^{2\phi}&=1\cr A_{0123}&=
-\left(1+{k_4\over r^4}\right)^{-1}\cr
ds^2&=\left(1+{k_4\over r^4}\right)^{-1/2}\eta_{\mu\nu}dx^{\mu}dx^{\nu}
+\left(1+{k_4\over r^4}\right)^{1/2}\left(dr^2+r^2d\Omega_5^2\right)\cr}}
where $\mu,\nu=0,1,2,3$ and $r$ is the radial coordinate for the
six-dimensional
transverse space. Note
that, being self-dual, the threebrane couples to the canonical metric.
We see immediately from \tbrane\ that there is no curvature singularity since
the radius of $S^5$ tends to the finite value $k_4^{1/4}$ as $r\to 0$. In
fact, the curvature scalar vanishes as a consequence of the self-duality.

Let us now consider the trajectory of a test threebrane falling radially
into a source threebrane, oriented along $x^a=\xi^a$ ($a=1,2,3$). Let the test
threebrane be either parallel or antiparallel to the source threebrane.
Although $F_{0123m}$ is, by anti self-duality \asdual, not the only
non-vanishing component of $F_{MNPQR}$, it is the only component
contributing to the equation of motion. Hence \tmotion\ yields
\eqn\tfbfb{-T_4 e^{4A}
\left({\dot t\over \sqrt{\dot t^2-\dot r^2 e^{-4A}}}\mp 1\right)=
-T_4(E\mp 1).}
The proper time is
\eqn\tblinel{d\tau^2=-ds^2=e^{2A}dt^2-e^{-2A}dr^2.}
The calculation now proceeds along the same lines as section 3. Again
the test threebrane takes a finite time to reach the source threebrane
and $dr/d\tau$ and $d^2r/d\tau^2$ both blow up at $r=0$.
Again in the antiparallel case with zero initial velocity
$\tau_0\sim k_4^{-1/2}\sim M_4^{-1/2}$.
The difference, of course, is that there never was
a curvature singularity to begin with!

\newsec{$D=11$ Supermembrane}

So far we have focussed our attention on solutions of $D=10$ supergravity,
the field theory limit of the superstring (and, presumably, of the
superfivebrane and superthreebrane). In \dfst, however, Duff and Stelle
found a supermembrane solution of $D=11$ supergravity. Indeed, the Dabholkar
{\it et al.} superstring solution in $D=10$ may be seen to follow from
the $D=11$ supermembrane solution by simultaneous dimensional reduction.
Let us denote all $D=11$ variables by a carat, and then make the ten-one split
\eqn\split{\hat X^{\hat M}=(X^M,X^2)\qquad\qquad\ M=0,1,3,...10.}
Then the solution \dabsol\ follows from
\eqn\redux{\eqalign{\hat g_{MN}&=e^{-\phi/6}g_{MN}({\rm canonical})\cr
\hat g_{22}&=e^{4\phi/3}\cr \hat A_{012}&=B_{01}.\cr}}
It was remarked upon in \refs{\dfluone,\dflutwo}\ that $\hat g_{MN}$
in \redux\ bears the same relation to $g_{MN}({\rm canonical})$ as
does the fivebrane $\sigma$-model metric in \fcansig. Indeed,
the $D=11$ metric $\hat g_{\hat M \hat N}$ is given
precisely by \dabfb\ with $\mu,\nu=0,1,2$. As such it is also manifestly
free of curvature singularities!

Just as for the self-dual threebrane, however, a test membrane falls
into a source membrane in a finite proper time.

\newsec{Black Strings and p-branes}

The situation for superstrings and super p-branes described thus far
differs radically from the non-supersymmetric string and p-branes, discussed
by Horowitz and Strominger\host. These display event horizons and also
singularities, even when the metric is written in the dual variables.
(For the self-dual black threebrane, the curvature scalar still vanishes
but $R_{MN}R^{MN}$ and $R_{MNPQ}R^{MNPQ}$ blow up). If we repeat the analysis
of sections 3 and 4 for these objects, we find that the test probe
always reaches the singularity in a finite proper time. Hence the
singularities are always real.

\newsec{Conclusion}

We have seen that, as far as singularities are concerned, the superstring
and the superfivebrane solitons are on an equal footing: the fivebrane is
a singular solution of fivebrane theory but a non-singular solution of
string theory while the string is a singular solution of string theory
but a non-singular solution of fivebrane theory. What is asymmetric,
however, is the state of current technology. One can prove rigorously
that $S({\rm string})$ of \smf\ describes the field theory limit of string
theory and that the string loop coupling constant is, from \scansig\
given by $g({\rm string})=e^{\phi_0}$; one has only plausibility arguments
that the dual action $S({\rm fivebrane})$ describes the field-theory limit
of fivebrane theory and that the fivebrane loop coupling constant is,
from \fcansig, $g({\rm fivebrane})=e^{-\phi_0/3}$ and hence that the strong
coupling limit of the string corresponds to the weakly coupled fivebrane and
vice-versa\dflutwo. Moreover, whereas the fivebrane solution can be
shown to be an exact solution of string theory to all orders in
$\alpha'=1/2\pi T_2$ using the methods of conformal field theory\chsone,
one can only conjecture that the string solution can be shown to be an
exact solution of fivebrane theory to all orders in $\beta'=1/(2\pi)^3 T_6$,
using some braney generalization of CFT.

\vfil\eject
\listrefs
\bye